\documentclass[12pt,aps,prb,preprint]{revtex4}   % style for Physical Review B and AJP are similar

\usepackage{amsmath}    % need for subequations
\usepackage{graphicx}   % for figures

\newcommand{\ajp}{AJP}  % example of a definition of a macro
\begin{document}

\title{Comment on `A close examination of the motion of an adiabatic piston' by Eric A. Gislason
[Am. J. Phys. \textbf{78}, 995--1001 (2010)]}
\author{Rodrigo de Abreu}
% \affiliation{Clark University, Department of Physics, Worcester,
%MA 01610}
\author{Vasco Guerra}
% \altaffiliation[Also at ]{Instituto de Plasmas e Fus\~ao Nuclear, IST, Lisboa, Portugal}  %  optional
 \email{vguerra@ist.utl.pt}   %optional
\affiliation{Departamento de F\'{\i}sica, Instituto Superior T\'ecnico, 1049-001 Lisboa, Portugal}
\date{\today}

\begin{abstract}
A recent paper by Gislason published in \ajp $\ $deals with the celebrated example of the so-called
``adiabatic piston'', a system involving two ideal gases contained in a horizontal cylinder and separated
by an insulating piston that moves without friction.\cite{Gislason2010}  
While the analysis presented in that paper is rather comprehensive, very interesting and useful as 
a teaching tool, it can be somewhat misleading if not taken within its appropriate context. As a matter of fact,
the evolution to equilibrium involves two
phases, a faster one leading to the equalization of pressures, and a slower one bringing the system to identical
temperatures. Although Gislason addresses only the first process, 
we note that the final state after the second phase, the evolution
to equal temperatures once the pressures are the same, \textit{is described by thermodynamics}. 
Therefore, the discussion of the adiabatic piston given by Gislason can and should be 
enriched, in order to promote a proper and general view of thermodynamics.
\end{abstract}

\maketitle

\section{Introduction}
In a recent paper published in \ajp, Eric A. Gislason puts forward a detailed analysis of the motion
of the ``adiabatic piston'' problem, consisting of two subsystems of the same ideal gas contained in a 
horizontal cylinder with insulating walls. \cite{Gislason2010}
Gislason  makes several important points and elaborates on the \textit{first process}, which brings the piston
to rest when the pressures of the two gases will be equal. Significant physical insight given by Gislason is the
damping of the piston motion as a result of the \textit{dynamic pressure} on the piston, ``because the pressure is
greater when the piston is moving towards the gas than when the piston is moving away from 
the gas.'' \cite{Gislason2010}
%\cite{Gislason2010,Abreu1996} 
Gislason cites several authors who have pointed out that 
``temperature and pressure fluctuations in the two gases will slowly act to bring the two temperatures to
equality.''\cite{Gislason2010} He correctly states that the ``time scale for this slow process is much longer than the 
time scale for the piston to come to rest''\cite{Gislason2010} and warns the reader that this 
slow process is not discussed in the paper.
Gislason asserts that ``thermodynamics cannot predict what the final temperatures will 
be'',\cite{Gislason2010} which is right only within the framework of the analysis of the first process.
%
%therefore being 
%unable to describe this \textit{second slow process}, 
Moreover, he adds that ``to achieve complete equilibrium the piston
must be able to conduct energy, which cannot occur for an adiabatic piston''.\cite{Gislason2010} The latter assertion
cannot be read in the context of the study of the faster process and, as detailed in the next section, 
is not rigorously valid if one keeps in mind the second process as well.
Notice that we find it very interesting to present
the analysis of the first process as done by Gislason, simply students should be aware of the approximations involved
and the conceptual problems it hides in some measure. The purpose of this comment is to contribute to
clarify this issue, by using the formalism of thermodynamics to extend the investigation to the second process as well.

An affirmation on the impossibility of thermodynamics to predict the final
equality of temperatures is a critical step, since a kinetic-statistical interpretation indeed
foresees the motion of the piston until the two temperatures are equal. In this line, an intuitive and beautiful 
discussion of this
second process is made
by Feynman,\cite{Feyetal1979} whereas a quantitative molecular dynamic simulation, establishing
the phenomenon beyond doubt, was published by
Kestemont and co-workers.\cite{Kesetal2000} 
The message we want to convey here
is that an accurate and careful use of thermodynamics must give the same 
final results as any kinetic simulation, as the latter is a microscopic 
interpretation of the results of the former.
In passing, let us announce we find that no occasion is too much to pay a tribute to the genius 
of Ludwig Boltzmann, and this is a perfect occasion to do so.

The remaining of this comment is structured as follows.
The way in which thermodynamics may handle the ``adiabatic piston''
problem is shown in the next section. A
short discussion and an identification of the origin of some common misunderstandings is given in section 
\ref{discussion}. Finally, section \ref{conclusions} summarises our main conclusions.

\section{Thermodynamics' answer to the problem}

Let us start by giving a straight answer to the ``adiabatic piston'' problem, namely, what are the final
pressures and temperatures of both gases, leaving further discussion for the
next section. 
The equality of pressures is a necessary condition, usually referred to as the condition for 
\textit{mechanical equilibrium}, corresponding to the first process. 
However, it is not sufficient for the complete equilibrium, the
\textit{thermodynamical equilibrium}, correlated to the second, slower, phase. 

It is worth noting that we cannot impose $dS=0$ once the 
pressures are equal,\cite{Abreu2002} although this is sometimes confused with the 
``adiabatic'' condition (\textit{cf.} next section). 
Indeed, there are configurations in the vicinity of the mechanical equilibrium, 
with greater global entropy, and the system will move towards these configurations.
Take note that the two subsystems are connected through the conditions of 
constant total volume and total energy. The collisions between the gas particles and the piston
will make the piston jiggle, \textit{allowing an exchange of energy} between both gases.
This energy exchanges will take place \textit{even if the piston is not a thermal conductor}, as they are
simply a result of the momentum transfer in the collisions (see the discussion by Feynman\cite{Feyetal1979}).
As a consequence, the system will indeed access the different available microscopic configurations
and move as a result of a blind entropic process, in accordance with Boltzmann's basic ideas 
and his microscopic interpretation of entropy.\cite{Boltzmann1872,Boltzmann1877} This is why the assertion that ``to achieve complete equilibrium,
the piston must be able to conduct energy, which cannot occur for an adiabatic piston'' does not hold.

Taking into account the preliminary discussion above, the system is described by the following set of 
equations\cite{Abreu2002}
\begin{eqnarray}
dU_1 = -p_1\;dV_1 + T_1\; dS_1 \label{dU1}\\
dU_2 = -p_2\;dV_2 + T_2\; dS_2 \label{dU2}
\end{eqnarray}
%Here, the following remark must be made. If the system is not in  
%equilibrium, the only possible interpretation
%of $p$ and $T$ is 
%\begin{eqnarray}
%p=-\left(\frac{\partial U}{\partial V}\right)_T\label{p}\\
%T=\left(\frac{\partial U}{\partial S}\right)_V 
%\end{eqnarray}
%Any other interpretation, such as the identification
%of these quantities with measurable pressure and temperature, and their relation with
%work and heat, risks to entail a misfortunate use of language and can induce unsound interpretations, as detailed in section \ref{discussion}.
Moreover, we have the condition
\begin{equation}
dS = dS_1 + dS_2\geq 0\label{dS}\ ,
\end{equation}
where the equality holds for the final equilibrium.
Equations (\ref{dU1}) and (\ref{dU2}) can be written in the form
\begin{eqnarray}
dS_1 = \frac{dU_1}{T_1} + \frac{p_1}{T_1}dV_1\label{dS1}\\
dS_2 = \frac{dU_2}{T_2} + \frac{p_2}{T_2}dV_2 \label{dS2}
\end{eqnarray}

Now, the piston jiggles, but, as long as the system reaches mechanical equilibrium, 
\begin{equation}\label{dEk}
dE_k = - dU_1 - dU_2=0\ ,
\end{equation}
where $E_k$ is the kinetic energy of the piston. Furthermore,
\begin{equation}
dV = dV_1+dV_2 = 0\ .
\end{equation}
Hence, $dU_2 = - dU_1$ and $dV_2 = -dV_1$. Substituting (\ref{dS1}) and (\ref{dS2}) in 
the equilibrium condition (\ref{dS}), we finally get
\begin{equation}\label{dS0}
dS = \left(\frac{1}{T_1}-\frac{1}{T_2}\right)dU_1 + \left(\frac{p_1}{T_1}-\frac{p_2}{T_2}\right)dV_1 \equiv 0\ .
\end{equation}
Therefore, the solution to our problem is 
\begin{eqnarray}
p_1=p_2\\
T_1=T_2
\end{eqnarray}
and \textit{both the mechanical and the thermodynamical equilibria are obtained}. Thermodynamics can and does predict
the final variables.

\section{Discussion}\label{discussion}

In the previous section we have shown that thermodynamics predicts correctly the evolution of the system to a
state of equal pressures \textit{and} equal temperatures. The reason the contrary inaccurate statement is repeated
by many authors is related to a problem of language and a misconceived notion associated with the
word ``adiabatic''. Actually, if the piston is ``adiabatic'', an additional condition is often imposed, based on 
an erroneous physical intuition, specifically,
\begin{equation}\label{dUi}
dU_i = -p_i dV_i\ \ \ , \ \ \ i=1,2 \ .
\end{equation}
%This is equation (7) from Gislason paper.\cite{Gislason2010}
The argument is that, since the piston is ``adiabatic'', $dQ=0$. 
If this would be the case we would have, substituting (\ref{dUi}) in (\ref{dS0}),
\begin{equation}\label{dS10}
dS = -\left(\frac{1}{T_1}-\frac{1}{T_2}\right)p_1\;dV_1 + \left(\frac{p_1}{T_1}-\frac{p_2}{T_2}\right)dV_1 \equiv 0\ .
\end{equation}
This expression would then be valid if the mechanical equilibrium $p_1=p_2$ holds, without the need for equality of the temperatures. In fact, 
using $p_2=p_1$ in (\ref{dS10}),
\begin{displaymath}
dS = -\left(\frac{1}{T_1}-\frac{1}{T_2}\right)p_1\;dV_1 + \left(\frac{1}{T_1}-\frac{1}{T_2}\right)p_1\;dV_1\ ,
\end{displaymath}
which is identically zero, regardless of the values of $T_1$ and $T_2$.

The critical point here is to realise that the additional condition (\ref{dUi}) \textit{is extraneous to the formalism}.
When the designation ``adiabatic piston'' is used, it is meant a piston \textit{with zero heat conductivity}.
If the piston is held in place (for instance, if it is fixed to the box by screws), then there is no
heat transfer from one subsystem to the other. Even though, if the piston is released, both systems are coupled, therefore
interacting and exchanging energy, as explained in the previous section, and there is
``heat exchange''. Trying to use the common
language, we could say that \textit{a piston which is
``adiabatic'' when it is fixed, no longer is ``adiabatic'' when it can move freely}! The condition $dQ=0$ cannot be imposed, as it
comes from a faulty instinct still somewhat related with the idea of caloric (associating ``heat'' to a fluid, or at least
being mislead by the designations ``heat'' and ``thermal insulator'' or ``adiabatic''), and completely misses
the subtlety of the concept of heat.

It is not too difficult to show that equation (\ref{dUi}) is not general and cannot be demonstrated.\cite{Abreu2002}
The conservation of energy is expressed by the first part of equation (\ref{dEk}), $dE_k+dU_1+dU_2=0$. 
On the other hand,
the work done on the piston is
\begin{equation}
dW = dE_k = \left(p_1^\prime - p_2^\prime\right) dV_1\ ,
\end{equation}
where $p_1^\prime$ and $p_2^\prime$ are \textit{dynamic pressures} (notice that they are denoted by
$P_1$ and $P_2$ in Gislason's paper\cite{Gislason2010}), \textit{i.e.}, the pressures the gases exert on the moving piston.
Therefore,
\begin{equation}
dU_1+dU_2 = -\left( p_1^\prime - p_2^\prime\right) dV_1\ .
\end{equation}
This does \textit{not} imply that (\ref{dUi}) is valid, as it does not require with generality $dU_i = -p_i^\prime dV_i$, although
this can be a good approximation during the fast process.
Hence, even after the first phase, when pressures are equal but the temperatures are still different,
we must write 
\begin{equation}
dU_i = -p_i\; dV_i + T_i\; dS_i \neq -p_i\; dV_i\ ,
\end{equation}
and (\ref{dUi}) is not general.

After the attainment of mechanical equilibrium the piston has no kinetic energy and the evolution to the final equilibrium verifies $dU_1=-dU_2$,
\textit{i.e.}
\begin{equation}
-p_1\;dV_1 + T_1 dS_1 = + p_2\; dV_2 - T_2 dS_2 \ .
\end{equation} 
Since $p_1=p_2$ and $dV_1 = -dV_2$, it comes
\begin{equation}
T_1 dS_1 = -T_2 dS_2\ .
\end{equation}
Finally, if $T_1>T_2$, and taking into account (\ref{dS}), $dS_2>0$ and $dS_1<0$, although the global change of entropy is
positive.\cite{Abreu2002} Accordingly, the temperatures $T_2$ and $T_1$ will slowly raise and decrease, respectively, until
both temperatures become equal and full equilibrium is achieved.

\section{Conclusion}\label{conclusions}

A very nice paper published recently in \ajp ~raises several interesting points on thermodynamics 
using the example
of the ``adiabatic piston''.\cite{Gislason2010} As asserted in that paper, 
the results it obtains must be used exclusively as a 
description of the first process of evolution to equilibrium, leading to mechanical equilibrium. 
However, the slow evolution
to thermodynamical equilibrium is also well described within classical thermodynamics and the complete
equilibrium is in truth achieved, even if the piston is not a thermal conductor. We believe this example
can be extremely useful in classroom to illustrate the subtleties around the concept of ``heat'', which goes beyond
the first ideas leading to its introduction in Physics. Besides, it helps advancing a more general 
and proper view of thermodynamics,
providing as well a strong
link to the microscopic interpretation of entropy. 
Additional appreciation of the problem, 
including the analysis of the first phase and the damped oscillations of the piston, 
can be found in a paper by Mansour and co-workers\cite{Manetal2006} and
in some former work of R. de Abreu.\cite{Abreu1996,Abreu2002b} Further discussion
on the concepts of work and heat is also available from Gislason and Craig.\cite{GC2005}

\begin{acknowledgments}
The authors are indebted to Professor Eric A. Gislason for the very rewarding discussion and his suggestions ensuing the submission of this comment, which
contributed to increase the quality and clarity of the manuscript.
\end{acknowledgments}

%\bibliography{Thermo}{}
%\bibliographystyle{unsrt}

\end{document}